\documentstyle{article}
\input psfig.sty
\input epsfig.sty

\newcommand{\be}{\begin{equation}}
\newcommand{\ee}{\end{equation}}
\newcommand{\bea}{\begin{eqnarray}}
\newcommand{\eea}{\end{eqnarray}}
\def\Re{{\cal R \mskip-4mu \lower.1ex \hbox{\it e}\,}}
\def\Im{{\cal I \mskip-5mu \lower.1ex \hbox{\it m}\,}}

\def\etal{{\it et al.}}

\def\tev{\,{\ifmmode\mathrm {TeV}\else TeV\fi}}
\def\gev{\,{\ifmmode\mathrm {GeV}\else GeV\fi}}
\def\mev{\,{\ifmmode\mathrm {MeV}\else MeV\fi}}

\begin{document}
\title{The noncommutative QED threshold energy versus the optimum collision energy}
\author{Zheng-Mao Sheng\thanks{Corresponding author.E-mail:zmsheng@css.zju.edu.cn}, ~ Yongming Fu and Haibo Yu\\
\footnotesize \it Department of Physics, Zhejiang University,
Hangzhou 310027, China
 }
\maketitle

\begin{abstract}
M\"{o}ller  Scattering and Bhabha Scattering on noncommutative
space-time is restudied. It is shown that the noncommutative
correction of scattering cross sections is not monotonous
enhancement with the total energy of colliding electrons, there is
an optimum collision energy to get the greatest noncommutative
correction. Most surprisingly, there is a linear relation between
the noncommutative QED threshold energy and the optimum collision
energy.

\vspace{5mm} \leftline{PACS:11.10.Nx,12.60.-i, 12.20.-m}
\end{abstract}

\section{Introduction}
The interest in formulating field theories on noncommutative
spaces is relatively old\cite{Snyder47}.It has been revived
recently due to developments connected to string theories in which
the noncommutativity of space-time is an important characteristic
of D-Brane dynamics at low energy
limit\cite{Connes98,Douglas98,SW99}. Much attention on
noncommutative field theory has been attracted
\cite{Nappi03,Sheikh-Jabbari99,Martin99,Ishibashi00,Ya99,Petriello01,Carroll01,Seiberg00,Sheikh00,Ihab00}.
It is shown that field theories on noncommutative space-time are
well defined quantum theories\cite{Douglas98}. From the point of
view of string theory, we need urgently to set up one theory that
is in conformity with existing elementary particle theory and can
be examined within the attainable range of energy scale in
experiment at present or in near future. It was generally thought
that prediction of string theory and noncommutative effect can
only be examined at Plank energy scale or at Great Unification
energy scale. However, Witten \etal propose recently that the
effect of the string maybe appear at the energy scale of \tev ,
the threshold value of commutativity of space, as D-brane dynamic
main nature in low energy limit, might as well be thought that its
energy scale is also at scale of \tev, or not far from that at
least\cite{Witten96}. In other hand,the energy scale of \tev~  is
just typical running energy scale of collide machine of next
generation\cite{Telnov90}. However, till now only a little
attention has  been paid on whether noncommutative effects can be
examined\cite{Mathews00,Hewett01}in experiments. In this letter,
we restudy the M\"oller scattering and Bhabha scattering in
noncommutative quantum electrodynamics(NCQED) to establish the
relation between NCQED threshold energy and the most optimum
collision energy while noncommutative correction of scattering
cross section arrives at the maximum and to discuss the method to
determine the NCQED threshold energy. Both M\"oller and Bhabha
scattering are pure lepton process. Their Standard Model result is
in accord with QED, so we only need discuss these process in QED
even if at the energy scale of \tev.

\begin{figure}
\vspace*{-0.8cm} \centering \epsfxsize=4.4in \epsfysize=2.4in
 \epsfbox{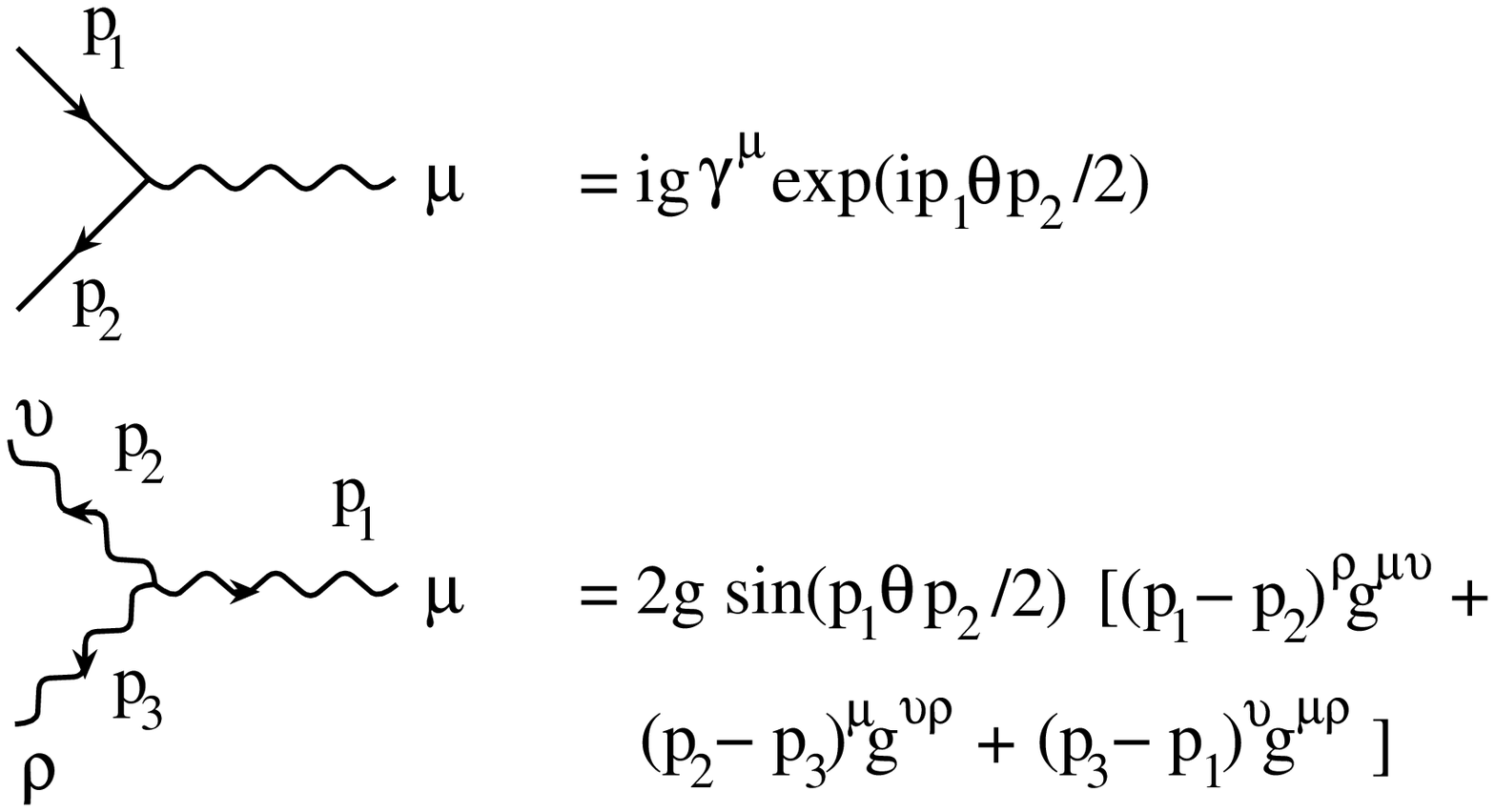}
\vspace*{-1.5cm}
 \epsfxsize=4.4in \epsfysize=2.8in
\epsfbox{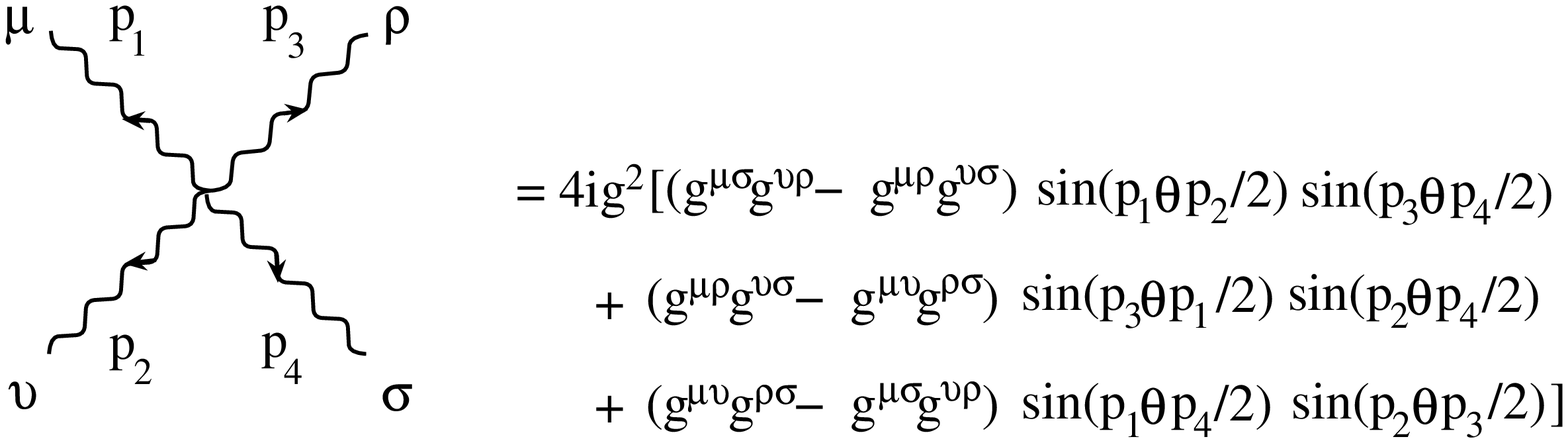} \vspace*{-1.5cm} \caption{Feynman rules for
interaction vertices of  NCQED } \label{fig1}
\end{figure}

The noncommutative space can be realized by the coordinate
operators satisfying
\be [\hat{X}_\mu,\hat{X}
_\nu]=i\theta_{\mu\nu}=\frac{1}{\Lambda_{NC}^2}i c_{\mu\nu}
\label{NCST}
\ee
 where $\theta_{\mu\nu}$ are the noncommutative
parameters, which is real, anti-symmetric and commutes with
space-time coordinate $\hat{X}_\mu$; $c_{\mu\nu}$ are
dimensionless parameters, and $\Lambda_{NC}$ denotes
noncommutative threshold energy . A noncommutative version of an
ordinary field theory can be obtained by replacing all ordinary
products with Moyal $\star$ products defined by
\begin{equation}
(f\star
g)(x)=exp\left(\frac{1}{2}\theta_{\mu\nu}\partial_{x^\mu}\partial_{y^\nu}\right)f(x)g(y)|_{y=x}
\label{StarP}
\end{equation}
we can get the NCQED Lagrangian by using above recipe as
\begin{equation}
{\cal L}=\frac{1}{2}i(\bar{\psi}\star \gamma^\mu D_\mu\psi
-(D_\mu\bar{\psi})\star \gamma^\mu \psi)- m\bar{\psi}\star
\psi-\frac{1}{4}F_{\mu\nu}\star F^{\mu\nu} \label{NCL}
\end{equation}
where
$D_\mu\psi=\partial_\mu\psi-ieA_\mu\star\psi$,$~~(D_\mu\bar{\psi})=\partial_\mu\bar{\psi}+ie\bar{\psi}\star
A_\mu$, $~~ F_{\mu\nu}=\partial_{\mu} A_{\nu}-\partial_{\nu}
A_{\mu}-ie(A_{\mu}\star A_{\nu}-A_{\nu}\star A_{\mu})$. This
Lagrangian can be used to obtain the Feynman rules for
perturbative calculations. The propagators for the free fermions
and gauge fields of NCQED are the same as that in the case of
ordinary QED. But for interaction terms, every interaction vertex
has a phase factor correction depending on  in and out 4-momentum.
Here we give out the interaction vertices of NCQED as that in
Fig.\ref{fig1}, where $p_1\theta p_2$ is the abbreviation for
$p_1^{\mu}\theta_{\mu\nu}p_2^{\nu}$.

\section{NCQED correction of cross section for M\"{o}ller Scattering}

\begin{figure}
\vspace*{-0.5cm} \centering \epsfxsize=4.0in \epsfysize=2.5in
 \epsfbox{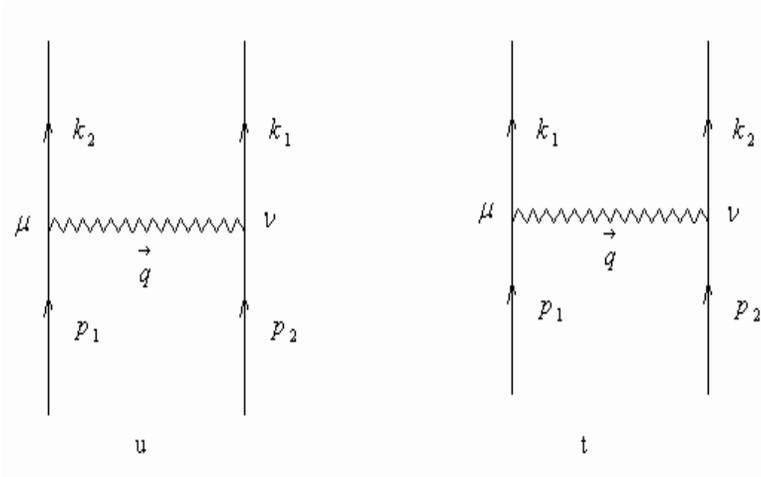}
 \caption{Feynman graph of
M\"{o}ller scattering } \label{fig2}
\end{figure}

The Feynman graphs of M\"{o}ller scattering considered the
contribution of the lowest rank tree graphs is shown as in
Fig.\ref{fig2}, which consists of u-channel and t-channel.
According to the  Feynman rules of NCQED, we can  get the
scattering amplitudes: \bea M_u &= -\bar
{u}^{r_2}(k_2)(-ig\gamma^\mu \exp(ip_1^{\rho}\theta_{\rho\sigma}
k_2^{\sigma}/2)u^{s_1}(p_1) \frac{-ig_{\mu\nu}}{q^2}\nonumber\\
& \cdot\bar {u}^{r_1}(k_1)(-ig\gamma^{\nu
}\exp(ip_2^{\rho}\theta_{\rho\sigma} k_1^{\sigma}/2)u^{s_2}(p_2),
\eea
\bea M_{t} &=\bar{u}^{r_1}(k_1)(-ig\gamma^\mu
\exp(ip_1^{\rho}\theta_{\rho\sigma}
k_{1}^{\sigma} /2)u^{s_1}(p_1)\frac{-ig_{\mu\nu}}{q^2}\nonumber \\
&  \cdot\bar {u}^{r_2}(k_2)(-ig\gamma^{\nu
}\exp(ip_2^{\rho}\theta_{\rho\sigma} k_2^{\sigma}/2)u^{s_2}(p_2).
\eea By the similar method used in the ordinary QED, we can obtain
the scattering section on noncommutative space at
ultra-relativistic limit as
\begin{equation}
\frac{d\sigma_{NC}}{d\Omega}=\frac{2\alpha^2}{s}
\left[\frac{8}{\sin^4\theta}-\frac{6}{\sin^2\theta}+{1\over2}
+\frac{2\cos(\phi_u-\phi_t)}{\sin^2\theta}\right]
\end{equation}
with Mandelstam variables
$$\phi_u={1\over2}
(p_1^{\mu}\theta_{\mu\nu} k_2^{\nu}+p_2^{\mu}\theta_{\mu\nu}
k_1^{\nu}),\,\,\, \phi_t={1\over2}(p_1^{\mu}\theta_{\mu\nu}
k_1^{\nu}+p_2^{\mu}\theta_{\mu\nu} k_2^{\nu}),$$ $
s=(p_1+p_2)^2=(k_1+k_2)^2.$ Let
$$\begin{array}{ll} p_1^{\mu}&= \frac{\sqrt{s}}{2}(1,-1,0,0)\nonumber \\
p_2^{\mu}&= \frac{\sqrt{s}}{2}(1,1,0,0),\nonumber \\
k_1^{\mu}&=\frac{\sqrt{s}}{2}\left(1,-\cos\theta,-\sin\theta\cos\phi,-\sin\theta\sin\phi\right)
\nonumber \\
k_2^{\mu}&=\frac{\sqrt{s}}{2}\left(1,\cos\theta,\sin\theta\cos\phi,\sin\theta\sin\phi\right)
\end{array}$$
\begin{equation}
 \phi_{M}\equiv\phi_u-\phi_t=\frac{s}{2\Lambda_{NC}^2}(c_{12}
\cos\phi-c_{31}\sin\phi)\sin\theta.\end{equation}

Comparing with the differential scattering cross section for
commutative case
\begin{equation} \frac{d\sigma_{C}}{d\Omega}=\frac{2
 \alpha^2}{s}\left[{8\over \sin^4\theta}-{6\over
 \sin^2\theta}+{1\over2}+{2\over\sin^2\theta}\right],\end{equation}
we get the correction for the differential scattering cross
section as
\begin{equation} \Delta\left(\frac{d
 \sigma}{d\Omega}\right)=\frac{2\alpha^2}{s}\left[{2\over
 \sin^2\theta}(1-\cos\phi_{M})\right]\label{DCS}
 \end{equation}
It is shown that $\Delta\left(\frac{d \sigma}{d\Omega}\right)$ is
not a monotonous function of $\sqrt{s}$, but appears some extremum
when $\phi_{M}$ is not very small. The maximum value of
$\Delta\left(\frac{d \sigma}{d\Omega}\right)$ appears when
$\sqrt{s}$ satisfy following relations
\begin{equation}
\cos(x)+x \sin(x)-1=0, ~{\rm and} ~~ \cos(x)<0 \label{CD}
\end{equation}
where $x=\left(\frac{\sqrt{s}}{\Lambda_{NC}}\right)^2\frac{(c_{12}
\cos\phi-c_{31}\sin\phi)\sin\theta}{2}$. The solutions of
eqs(\ref{CD}) is $x=2.33$,$~~9.21$,$~~15.58$,$~~21.9, ...$.

It means that the collision energy $\sqrt{s}$ corresponding to the
maximum $\Delta\left(\frac{d \sigma}{d\Omega}\right)$ is direct
proportion to the noncommutative threshold energy $\Lambda_{NC}$.
The minimum of the ratio is 2.16 when we take $c_{12}=1,~c_{31}=0$
and $\phi=0,~~\theta=\frac{\pi}{2}.$

Moreover, one can expect the maximum correction of the total
scattering cross section $\Delta\sigma$ will appear and the
optimum colliding energy will also be direct proportion to the
noncommutative threshold energy $\Lambda_{NC}$.

\begin{figure}
\centering \epsfxsize=4.0in \epsfysize=2.6in \epsfbox{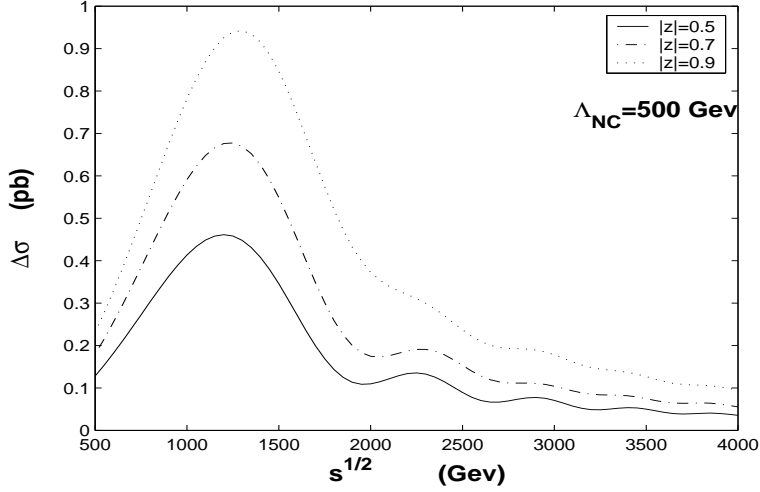}
\caption{The relation between $\Delta\sigma$ and $\sqrt{s}$ for
Moller scattering} \label{fig3}
\end{figure}

The correction curves for total scattering cross section
$\Delta\sigma$ vs colliding energy $\sqrt{s}$ as shown in
Fig.\ref{fig3} in which we take
$\Lambda_{NC}=500\gev$,$c_{12}=1,~c_{31}=0$ and an angular cut of
$ |\cos\theta| \le |z|=0.9(0.7,0.5) $.

\begin{figure}
\centering \epsfxsize=4.0in \epsfysize=2.6in \epsfbox{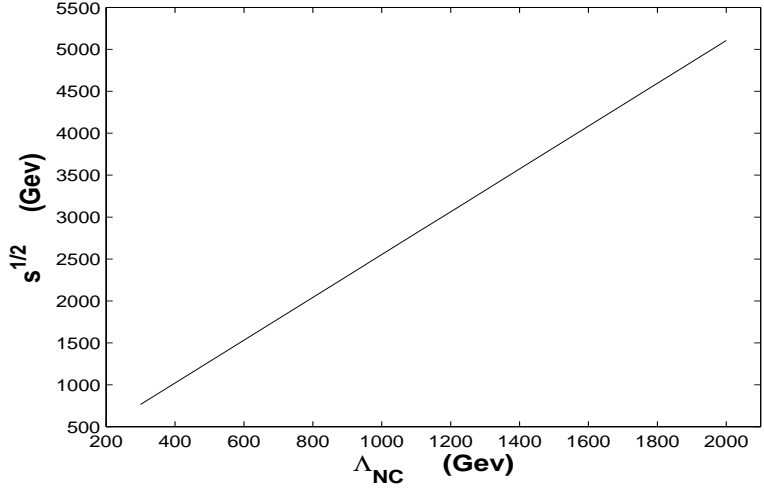}
\caption{The optimum collision energy $\sqrt{s_o}$ vs the NCQED
threshold energy $\Lambda_{NC}$ for M\"{o}ller scattering}
\label{fig4}
\end{figure}

From Fig.\ref{fig3}, we really find that $\Delta\sigma$ does not
always increase with $\sqrt{s}$, but exists as a kurtosis
distribution. Under the condition of $\Lambda_{NC}=500\gev$, a
maximum appears when the energy of collision particles $
\sqrt{s_o} =1288.8\gev$. This implies that there is an optimum
collision energy to observe the noncommutative effect in
M\"{o}ller scattering, if the NCQED threshold energy is fixed.
However, we do not know the exact value of the NCQED threshold
energy. We have to determine the NCQED threshold energy
$\Lambda_{NC}$ at the first. To do this, it is useful to establish
the relation between the NCQED threshold energy and optimum
collision energy $\sqrt{s_o}$, which is shown as in
Fig.\ref{fig4}.
After curve fitting, we get
\begin{equation} \sqrt{s_o}=0.000001+2.5776\Lambda_{NC}.\label{linear1}
\end{equation}
This means that the optimum collision energy is about two and half
times of the NCQED threshold energy. If we determine the optimum
collision energy $\sqrt{s_o}$ for noncommutative effect in the
M\"{o}ller scattering in next generation colliding experiment by
increasing gradually the energy of collision particles, we can
determine the NCQED threshold energy $\Lambda_{NC}$ by relation
eq.(\ref{linear1}).

\section{NCQED correction of cross section for Bhabha Scattering}

\begin{figure}
\centering \epsfxsize=4.0in \epsfysize=2.6in \epsfbox{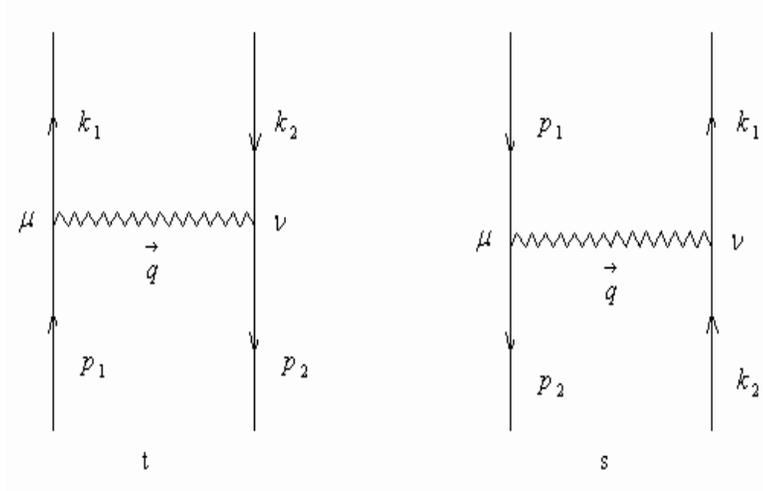}
\caption{The Feynman graphs of Bhabha scattering } \label{fig5}
\end{figure}

In order to know if the linear relation between the NCQED
threshold energy and the optimum collision energy  is universal,
we restudy Bhabha scattering. Its Feynman graphs with the lowest
rank tree graphs is shown as in Fig.\ref{fig5}, which consists of
t-channel and s-channel.
The scattering amplitudes of two graphs  are respectively \be
M_{t} = -\frac{ig^2\exp(i\phi_t)}{(p_1-k_1)^2}(\bar
{u}^{s_2}(k_1)\gamma^{\mu}u^{s_1}(p_1))\bar
{v}^{r_2}(p_2)\gamma_{\mu}v^{r_1}(k_2)\ee \be
M_{s}=\frac{ig^2\exp(i\phi_s)}{(p_1+p_2)^2}\left(\bar
v^{r_1}(p_2)\gamma^{\mu}u^{s_1}(p_1)\right)\bar
{u}^{s_2}(k_1)\gamma_{\mu}v^{r_2}(k_2) \ee where
\begin{equation}\phi_t={1\over2}(p_1^{\mu}\theta_{\mu\nu}
k_1^{\nu}-p_2^{\mu}\theta_{\mu\nu} k_2^{\nu}),\,\,\,\,\,
\phi_s={1\over2}(p_2^{\mu}\theta_{\mu\nu}
p_1^{\nu}-k_2^{\mu}\theta_{\mu\nu} k_1^{\nu}).
\end{equation}
The differential scattering cross section for Bhabha scattering on
noncommutative space at ultra-relativistic limit is
\begin{equation}\frac{d
\sigma_{NC}}{d\Omega}=\frac{\alpha^2}{4s}(3+\cos^2\theta+
2\frac{(1+\cos\theta)(4-\sin^2\theta\cos\phi_{b})}{(1-\cos\theta)^2}
)
\end{equation}
with  $ \phi_{b}\equiv\phi_s-\phi_t$.

To take
$$\begin{array}{ll} p_1^{\mu}  =
\frac{\sqrt{s}}{2}(1,-1,0,0),\\
k_1^{\mu}=\frac{\sqrt{s}}{2}\left(1,-\cos\theta,-\sin\theta\cos\phi,
-\sin\theta\sin\phi\right), \\
p_2^{\mu} =  \frac{\sqrt{s}}{2}(1,1,0,0),\\
k_2^{\mu}=\frac{\sqrt{s}}{2}\left(1,\cos\theta,\sin\theta
\cos\phi,\sin\theta\sin\phi\right),
\end{array}
 $$
then
\begin{equation} \phi_{b}=\frac{-s}
{2\Lambda_{NC}^2}[c_{01}(1-\cos\theta)-c_{02}
\sin\theta\cos\phi-c_{03}\sin\theta\sin\phi].
\end{equation}
Comparing with the differential scattering cross section of Bhabha
scattering in ordinary QED case
\begin{equation} \frac{d
\sigma_{C}}{d\Omega}=\frac{\alpha^2}{4s}\left[3+\cos^2\theta+
2\frac{(1+\cos\theta)(4-\sin^2\theta)}{(1-\cos\theta)^2}\right],
\end{equation}
we obtain the correction of differential scattering cross section
for noncommutative effect as
\begin{equation}
\Delta\left(\frac{d
 \sigma}{d\Omega}\right)_{B}=\frac{\alpha^2}{4s}\left[
 {2(1+\cos\theta)^2\over
 (1-\cos\theta)}(1-\cos\phi_{b})\right].
 \end{equation}
It is noted that the noncommutative effect increase the cross
section in the Bhabha scattering in opposition to the case in the
M\"{o}ller scattering. The maximum value of $\Delta\left(\frac{d
 \sigma}{d\Omega}\right)_{B}$  will also appear when
$\sqrt{s}$ satisfy following relations
\begin{equation}
\cos(x)+x \sin(x)-1=0, ~\rm{and} ~~\cos(x)<0 \label{CD2}
\end{equation}
where
$x=\left(\frac{\sqrt{s}}{\Lambda_{NC}}\right)^2\frac{c_{01}(1-\cos\theta)-c_{02}
\sin\theta\cos\phi-c_{03}\sin\theta\sin\phi}{2}$. Because the
eq.(\ref{CD2}) is the same as eq.(\ref{CD}) besides  the
difference of the expression of $x$ , the linear relation between
the NCQED threshold energy and the optimum collision energy will
still exist.

\begin{figure}
\centering \epsfxsize=4.0in \epsfysize=2.6in
\epsfbox{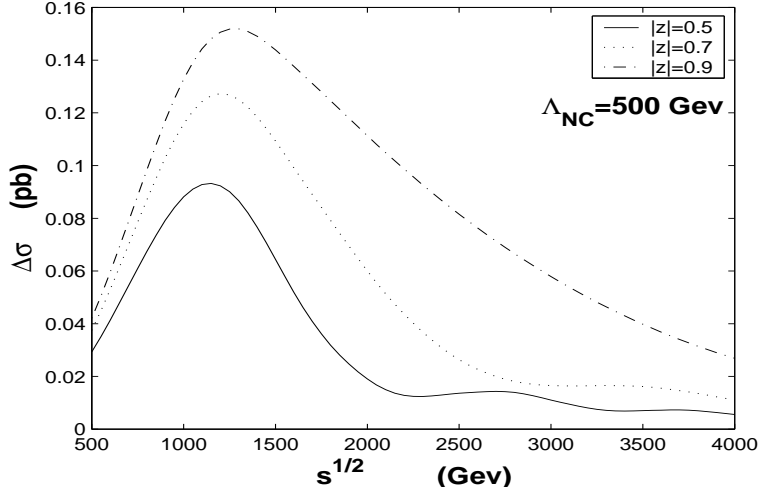} \caption{The relation between
$\Delta\sigma$ and $\sqrt{s}$ for Bhabha scattering with
$c_{01}=1,\,\, c_{02}=c_{03}=0$.} \label{fig6}
\end{figure}

The correction curves for total scattering cross section
$\Delta\sigma$ vs colliding energy $\sqrt{s}$ as shown in
Fig.\ref{fig6} in which we take $\Lambda_{NC}=500\gev$,
$c_{01}=1,\,\, c_{02}=c_{03}=0$ and an angular cut of $
|\cos\theta| \le |z|=0.9(0.7,0.5) $.

\begin{figure}
\centering \epsfxsize=4.0in \epsfysize=2.6in
\epsfbox{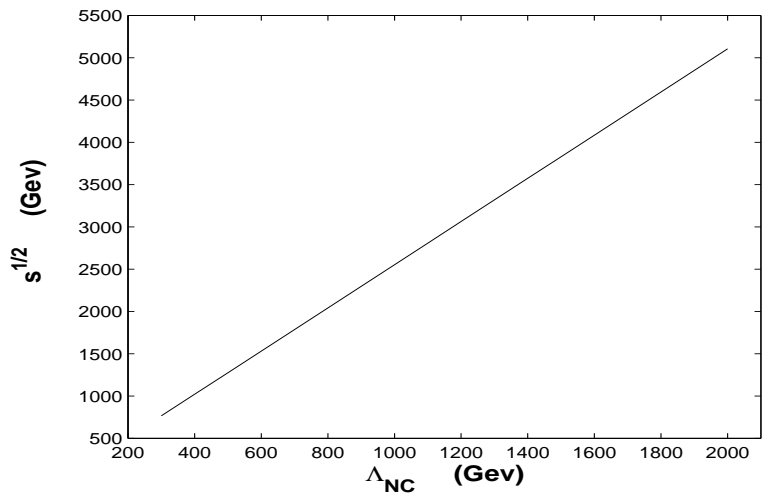} \caption{The relation between the optimal
collision energy $\sqrt{s_o}$ and noncommutative threshold energy
$\Lambda_{NC}$ for Bhabha scattering with $c_{01}=1,\,\,
c_{02}=c_{03}=0$ .} \label{fig7}
\end{figure}

From Fig.\ref{fig6}, we  also find that $\Delta\sigma$ does not
always increase with $\sqrt{s}$, but exist as a kurtosis
distribution. Under the condition of $\Lambda_{NC}=500\gev$, a
maximum appears when the energy of collision particles $
\sqrt{s_o} = 1276.5 \gev$. It show that there also is an optimum
collision energy to observe the noncommutative effect in Bhabha
scattering. The relation between the NCQED threshold energy and
optimum collision energy is shown as in Fig.\ref{fig7}.

After curve fitting, we obtain
\begin{equation}
\sqrt{s_o} =-0.0035+2.5529\Lambda_{NC}\hspace{2cm} \label{linear2}
\end{equation}
This means that there really is a linear relation between the
optimum collision energy and the NCQED threshold energy
$\Lambda_{NC}$. Moreover, its slope is almost the same as that in
M\"{o}ller scattering. If we determine the optimum collision
energy for noncommutative effect in the Bhabha or M\"{o}ller
scattering in next generation colliding experiment, we can
determine the NCQED threshold energy $\Lambda_{NC}$ by relations
eqs.(\ref{linear1},\ref{linear2}). In other hand, we shall choose
the optimum collision energy as the running energy of the next
generation collider to explore the noncommutative effect.

\section{Summary and discussions}

Firstly, we find that it is not the easier to explore
noncommutative effect, the higher collision energy in both
M\"{o}ller and Bhabha scattering, but there is an optimum
collision energy to get the greatest noncommutative correction for
total scattering cross section. Secondly, we find the linear
relations eqs.(\ref{linear1},\ref{linear2}) between the optimum
collision energy and the NCQED threshold energy$\Lambda_{NC}$, the
ratio is about 2.5 for both M\"{o}ller and Bhabha scattering. If
the optimum collision energy is determined by increasing gradually
the energy of collision particles from Hundreds \gev \ to several
\tev \ in the next generation collider, then the NCQED threshold
energy can be determined from the above linear relations. The
ordinary QED scattering cross section is inverse proportion to
square of the collision energy $s$, the NCQED correction of the
cross section is direct proportion to $[1-\cos(\Delta\phi_{NC})]$
where $\Delta\phi_{NC}$ is the noncommutative phase factor
correction which is direct proportion to $s$, so we argue that the
linear relation between $\sqrt{s_O}$ and $\Lambda_{NC}$ will be
satisfied by most QED scattering process. It will be the most
efficiency to explore the noncommutative effects by choosing the
optimum collision energy $\sqrt{s_O}$ as the operating energy of
the colliding experiment.

Authors thank Dr.H.X. Yang, Profs.C.R. Nappi, M.X.Luo and D.H.Lu
for useful discussion. The work is supported partly by funds from
Pandeng Project of China, NSFC (Grant NO. 90303003) and Zhejiang
Provincial Natural Science Foundation of China.


\end{document}